\documentclass[journal=jacsat,manuscript=article]{achemso}
\usepackage{amsfonts,amsthm,amsmath,amssymb}
\usepackage{url,hyperref}
\usepackage{graphicx}

\author{Anna Lappala}
\author{Saahil Mendiratta}
\author{Eugene M. Terentjev}
\email{emt1000@cam.ac.uk}
\affiliation{Cavendish Laboratory, University of Cambridge, JJ Thomson Avenue, Cambridge CB3 0HE, U.K.}

\title{Arrested spinodal decomposition in polymer brush collapsing in poor solvent}
\begin{document}

\begin{abstract}
\noindent
We study the Brownian dynamics of flexible and semiflexible polymer chains densely grafted on a flat substrate, upon rapid quenching of the system when the quality of solvent becomes poor and chains attempt collapse into a globular state. The collapse process of such a polymer brush differs from individual chains, both in its kinetics and its structural morphology. We find that the resulting collapsed brush does not form a homogeneous dense layer, in spite of all chain monomers equally attracting each other via a model Lennard-Jones potential. Instead, a very distinct inhomogeneous density distribution in the plane forms, with a characteristic length scale dependent on the quenching depth (or equivalently, the strength of monomer attraction) and the geometric parameters of the brush. This structure is identical to the spinodal-decomposition structure, however, due to the grafting constraint we find no subsequent coarsening: the established random bundling with characteristic periodicity remains as the apparently equilibrium structure. We compare this finding with a recent field-theoretical model of bundling in a semiflexible polymer brush.
\end{abstract}


\section{Introduction}\label{section1}
The theoretical problem of polymer chain collapse in poor solvents has received
much attention over the past decades, both from analytic~\cite{lifshitz_1969,moore_1976,kholodenko_1984a,muthukumar_1984,descloiseaux_1990} and numerical approaches~\cite{baumgartner_1987,kremer_1988}. This generic collapse, often called the ``coil-globule transition'', is well-understood as a local version of the polymer demixing Flory has originally formulated for polymer solutions~\cite{flory_1942}. The coil-globule transition is closely associated with the initial stages of protein folding~\cite{banavar_2004}. Collapse takes place as a result of effective pair interaction between chain monomers becoming attractive, e.g. due to changing solvent quality. Recent studies of the collapse dynamics have shown that it is a non-trivial kinetic process, with competing interactions leading to inhomogeneous intermediate structures forming and persisting on the way to the final globular state~\cite{degennes_1985,halperin_2000,lappala_2013}.

In this work, we look at the collective behaviour of several chains in a confined configuration known as ``polymer brush''. By definition, a polymer brush is a layer of polymers attached with one end to a surface at a sufficiently high grafting density (i.e. a sufficiently close distance between individual polymer chains). In practice, the attachment can take place via a covalent bond, a specific ligand, physical adsorption or self-assembly. In our simulations, the chains have their first monomer fixed on a flat surface, resulting in a constant grafting density throughout the simulation. Chains themselves can vary in length, composition and individual properties, from very primitive ideal chains to complex models incorporating block copolymers, electric charge and hydrodynamics.  The subclass of charged (polyelectrolyte) brushes is even more interesting in biological context.  There are numerous examples of polymer brushes in biological systems such as protein micelles, planar brushes of polysaccharides, cylindrical polymer brushes in microtubules and neurofilaments.  In terms of useful physical, chemical and biomedical applications, brushes can be used to stabilise colloids, lubricate surfaces, deliver drugs by biodegradable micelles, in DNA microarrays for diagnostic analysis of mutations and to reduce friction in artificial joints, to name but a few.~\cite{ayres_2010} Here we present a simple, planar  brush of self-avoiding bead-and-spring polymer chains without additional electrostatic interactions or hydrodynamics, which makes the analysis tractable and comparable with well-established theories -- but we also examine the case of chains with a long persistence length as a crude attempt to account for polyelectrolyte stiffening. The simple picture of polymer brushes may serve as the basis for models in diverse interfacial systems in biophysics and polymer science, such as polymeric surfactants, stabilised suspensions of colloid particles, and many structures formed by block copolymers.

The recent theory of Benetatos et al.~\cite{benetatos_2013} considered stiff and semiflexible chains in a brush configuration, and  predicted a bundling instability when an attractive interaction between monomers exceeds a critical value.  In their work, Benetatos et al.  established that there is a competition between permanent grafting favouring a homogeneous state of the polymer brush and the non-localised attraction, which tends to induce in-plane collapse of the aligned polymers, giving rise to an instability of the homogeneous phase to a bundled state. In the bundled state, the density in the plane is modulated by a length scale that depends on the strength of the attractive interaction.

In this work we carry out coarse-grained Brownian dynamics simulations of a collection of chains uniformly grafted to a flat repulsive surface, to find their equilibrium swollen conformation and then follow their behavior after instant quenching. We use the parameters setup of Rosa and Everaers~\cite{rosa_structure_2008} within the Large Scale Atomic/Molecular Massively Parallel Simulator (LAMMPS) package~\cite{steve_plimpton_fast_1995}.  The initial structure of the polymer chains was configured to be linear, and the first step was to equilibrate the linear brush into a constricted coil structure in good solvent. Due to the high density of grafting, in a good solvent their individual conformation is quite stretched out, and we find a good agreement with classical results on the parabolic density profile~\cite{milner_1988}. After equilibration, the solvent conditions were instantly switched to poor solvent, resulting in chains collapsed onto their flat substrate surface. Here we find an unexpected effect of spinodal decomposition in the plane of collapsing brush, and study it quantitatively through density profile-, potential energy-  and characteristic length-scale measurements, as well as the kinetics of collapse. In all cases, we compared the behaviour of flexible and semiflexible polymer brushes upon collapse over time.

Brownian dynamics simulations allowed us to follow the density evolution of brushes over time, as it evolves from the equilibrium parabolic profile in the swollen brush to the densely packed layer on the surface. The characteristic length scale of bundled regions in the collapsed brush appear to agree with the theoretical prediction of the square root dependence of this length scale on the attractive potential well depth~\cite{benetatos_2013}.

\begin{figure}[t]
\centering
\includegraphics[width=.75\textwidth]{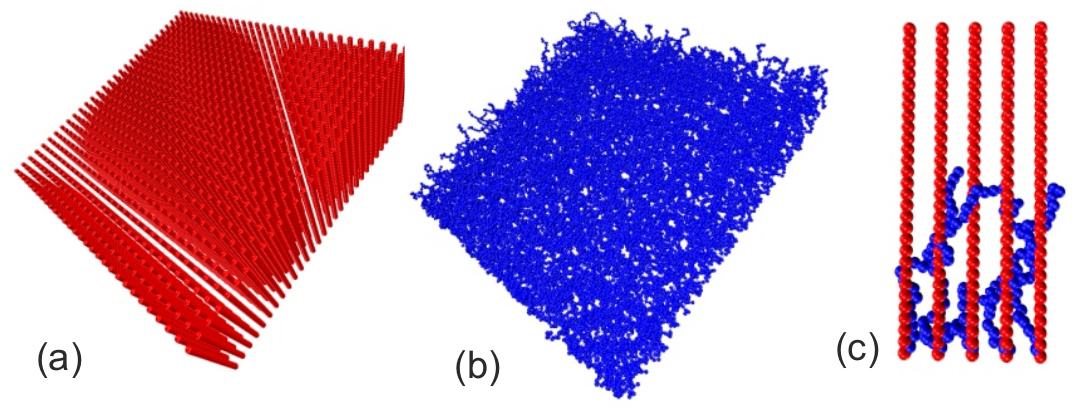}
\caption{ (a) The initial configuration of our simulation, illustrating the grafting structure and density. In this case, the grafting on the $x$-$y$ plane was on a uniform square lattice with the spacing $a=3\sigma$  (b) The equilibrated brush in a good solvent. (c) Comparison of the initial structure and the equilibrated coil. }
\label{fig1}
\end{figure}


\section{Simulation of polymer brush} \label{section2}

The model used for numerical simulations in this work is based on the bead-spring model described in the molecular dynamics context by Kremer and Grest~\cite{kremer_dynamics_1990} and Plimpton~\cite{plimpton_2003}.  We take an individual polymer chain composed of connected monomeric units consisting of $N=30$ monomers, and 900 of these chains formed a brush on a 30x30 square lattice as shown in fig.~\ref{fig1}. Model parameters are similar to the ones described by Lappala et al.\cite{lappala_2013} :  each monomer has a diameter of $\sigma$ in reduced units, which is chosen to be 0.3 nm in real units (a typical size of an amino acid residue in proteins), $\kappa$ and $R_0$ of the FENE potential were set so as to avoid any bond crossing. For our simulations, the values were used as in ref.\cite{kremer_dynamics_1990} : the maximum bond length $R_0= 1.5\sigma$ and the spring constant $\kappa=30\epsilon / \sigma^2$. The spring constant was specifically studied by Kremer and Grest~\cite{kremer_dynamics_1990} and was found to be strong enough such that the maximum extension of the bond was always less than 1.2$\sigma$ for $k_{\mathrm{B}}T= 1.0\epsilon$, making the covalent bond breaking energetically unfeasible. The important ratio measuring the local bending modulus of the chain, $ K_\theta$/$k_{\mathrm{B}}T$  (the chain persistence length $l_p = \sigma K_\theta / k_{\mathrm{B}}T$) was set to  $l_p = \sigma$ (or Kuhn's length of 0.6 nm) for a case of flexible chains, and to  $l_p = 8\sigma$ for a semiflexible brush with the Kuhn's length of 4.8 nm. Integration time step was chosen to be $\Delta t_\mathrm{int}=0.01\tau$ where $\tau$ is the reduced (Lennard-Jones) time, defined as $\tau=\sigma(m/\epsilon)^{0.5}$ (that is, determined by the ratio of particle mass and the depth of the Lennard-Jones (LJ) potential, $\epsilon$) as discussed in refs.\cite{kremer_dynamics_1990, rosa_structure_2008}. Taking the value of $\sigma$ above, $\epsilon =k_{\mathrm{B}}T$, and the typical mass of an aminoacid residue (molecular weight $\approx 100$), we obtain the estimate of $\tau \approx 1.96$ps. Hence, with $10^{7}$ time-steps  of this coarse-grained simulation we are able to follow the dynamics of a polymer brushes for 196 nanoseconds. Interactions between monomers are described by the standard potentials: FENE for connected pairs of particles, bending elasticity
$K_\theta (1-\cos \theta) $
with $\theta$ the angle formed between two consecutive bonds, and the LJ potential representing a longer-range interaction between any two particles. In good solvent the LJ potential was set to zero at $r_\textrm{cut-off} = 2^{1/6}\sigma $, retaining only the soft repulsion between monomers.
In order to simulate poor solvent conditions when there is an effective attraction between monomers, the cut-off value was set to $r_\textrm{cut-off} =3 \sigma$, while retaining the depth of $\epsilon$.

The initial parameters of the set up play an important role in the physics of the simulated brush collapse, and therefore the parameters used to build a brush had to be chosen carefully. Considering a box in the Cartesian frame, the brush was created using a $30 \times 30$ square plane lattice at $z = 0$, with lattice spacing of $3\sigma$. The length of the brush has a direct impact on the grafting density, and hence the optimal length of the brush was defined to be $N=30$ -- if the chains were longer, the 900 chains would make the simulation much heavier and acquisition of the required statistics more difficult. The lattice spacing (or grafting density) of $3\sigma$ had to be chosen such that it results in ample interactions between chains in good solvent -- if this number was higher, the chains would simply behave as individual coils, also described as ``mushrooms''~\cite{milner_1988} with few inter-chain interactions.
The grafting density was chosen on the basis of the cut-off radius used for the attractive potential, and also based on the  volume of a collapsed globule, which for an $N=30$ chain should be $30\cdot(4/3)\pi\cdot(\sigma)^3$ giving a linear distance of $\sim 5\sigma$, so we are assured of the overlap in the collapsed brush.

The first atom of each chain was bound to the $z = 0$ plane which was given a repulsive Lennard-Jones potential to keep the chains restricted to the positive $z$ direction. In order to confirm that the $N=30$ chains of a polymer brush were initially equilibrated in a good solvent, we simulated a single grafted chain of 1001 atoms, relaxing from an initial straight configuration to a random self-avoiding coil. This equilibrium state was confirmed visually and by density measurements for the standard random coil.  It is important that commonly used measurements of potential energy relaxation as a function of time are not suitable for judging equilibration, because the potential energy of a chain quickly reaches a very stable plateau for a `crumbled' extended chain conformation, which then takes a very long time to equilibrate into a proper self-avoiding coil. After the equilibrium of a long chain was confirmed after 10 million timesteps, we used this time for a polymer brush with much shorter chains on a planar surface to assure ourselves that it was properly equilibrated.

Once an equilibrium of grafted chains was achieved, an attractive LJ potential between all particles (poor solvent conditions) was switched on and the coil collapsed into a globule. Among other parameters, we monitored the collapse by the average areal density.
Density of the brush $\rho (z)$ can be described from the condition that an integral $\int_0^\infty \! \rho(z)\,\mathrm{d}z$ is equal to the total number of particles, which in our simulations is $900\times30=27000$.
We calculated the density along the $z$-direction, at each time step. The entire box was divided into parallel bins (in $xy$-plane) of unit thickness along the $z$-direction. The total number of particles in each bin forms a one-dimensional array $\rho(z)$, which we present as the non-normalised density.
This gives a constant density profile at $\rho(z)=900$ for $z\leq 30$ for the initial straight configuration of chains. For averaging of equilibrium structures, each $z$-bin was averaged over 5 brush configurations separated by 500,000 ts. According to the classical theory of expanded brush~\cite{milner_1988}, in a good solvent the density $\rho(z)$ will have a parabolic profile, which we accurately record in our simulations (see fig.~\ref{fig5} below). In the dynamic process of collapse we has to present an instantaneous $\rho(z)$ without time-averaging, resulting in a certain noise in corresponding profiles.

\section{Collapse of polymer brush}\label{section3}

In order to simulate the collapse of a polymer brush, the initial linear chains have to be fully equilibrated in conditions that imitate good solvent, which will result in a realistic simulation of collapse -- if the chains are not equilibrated in such a way, the collapse will happen by rolling of straight segments into a globular structure, which is not a realistic response. After equilibration, the solvent conditions are rapidly changed to poor -- a process known as quenching -- allowing the coils to collapse in an entangled, irregular manner. This more realistic collapse results in an unexpected highly inhomogeneous structure in the plane, fully resembling spinodal decomposition morphology. These structures were highly stable at equilibrium and their structural properties, such as the characteristic length scale of the formed collapsed structures, were dependent on the strength (the well-depth $\epsilon$) of the attractive potential. We followed the same principle as with verifying equilibration in an extended brush in good solvent. We first recorded the potential energy relaxation of the system until it reached an apparent plateau -- and then waited for another 26 million time-steps for further equilibration (it was only 10 Mts needed for the $N=1000$ chain to equilibrate into a coil). In contrast to the annealed case of polymer coils, here we have no structural criteria to verify equilibration; in fact, one might argue that in the dense collapsed mesh of mutually attracting chains the full equilibrium is unattainable on any reasonable time scales; although the system is not a glass, the internal mobility must be very low). However, we were certain that neither the morphological features of spinodal decomposition patterns, nor the average density profiles have not changed between $\sim 50$ kts and $26$ Mts. Hence we declare these states equilibrium with our accuracy.

\begin{figure} [t]
\centering
\includegraphics[width=.99\textwidth]{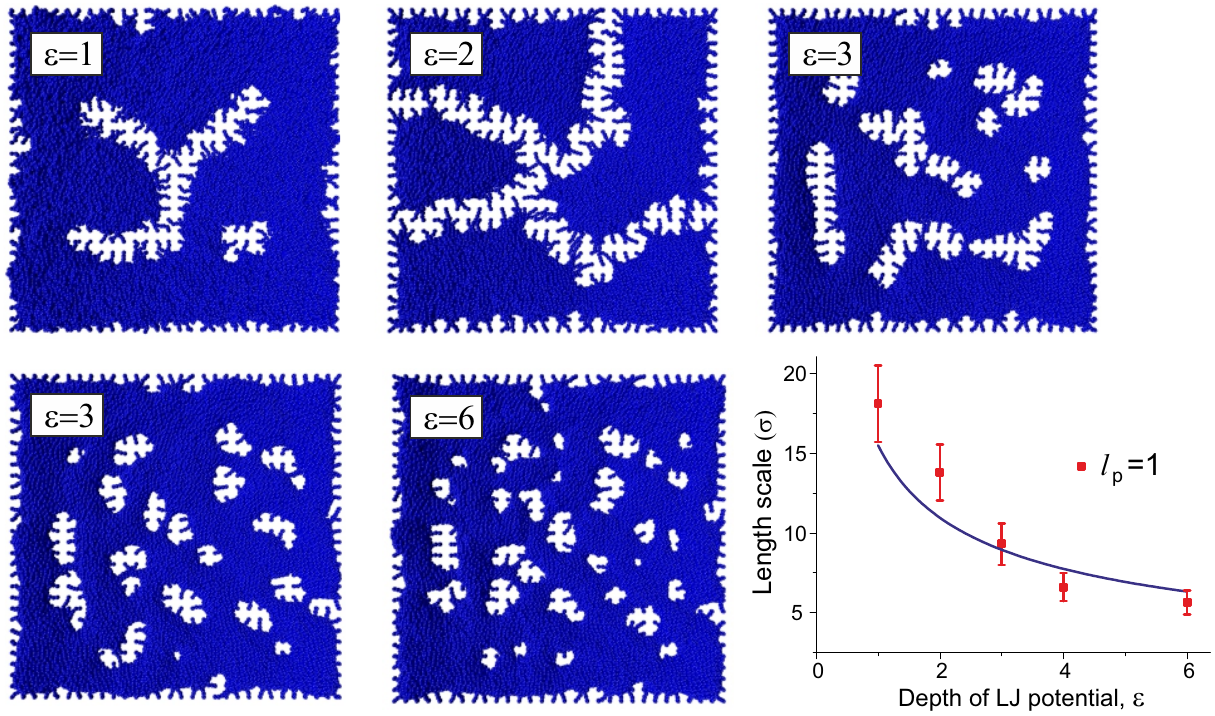}
\caption{Morphological properties of the collapsed flexible polymer brushes ($l_p=1$) at varying attractive potential well-depths, $\varepsilon$, labelled on each panel. The plot of the average characteristic length scale of the spinodal pattern is fitted to the square-root dependence: $\xi \sim 1/\sqrt{\varepsilon}$.  }
\label{fig2}
\end{figure}

As the polymer brush collapses, we see the initially homogeneous distribution of monomers in the $xy$-plane breaking down and a phenomenon that appears indistinguishable from the classical spinodal decomposition taking place. This clustering of polymer chains depends on the strength of the attractive potential between individual chains of the brush, characterised by a ratio $\varepsilon = \epsilon / k_\mathrm{B}T$. In flexible chains, we found distinctive coarsening and large clusters at small potential well depths, and increasingly smaller-sized clusters of polymers as the potential well depth increases, as shown in  fig.~\ref{fig2}.
Although the difference in characteristic length scale of the structure, $\xi$, is apparent between different panels, it was not easy for us to determine it quantitatively. Due to the relatively small area of the $xy$-map, the Fourier transformation methods (so useful in classical studies of spinodal decomposition, in Cahn-Hilliard model and beyond) were not possible in our case. In the end we have resorted to manually extracting many independent measurements of `thickness' of the dense-polymer regions (a distance perpendicular to both interfaces) and averaging the result. The final panel in fig.~\ref{fig2} presents such an average characteristic length scale, as a function of increasing strength of the attractive LJ potential, which we fitted to a square-root dependence $\xi \sim 1/\sqrt{\varepsilon}$.

\begin{figure} [t]
\centering
\includegraphics[width=.99\textwidth]{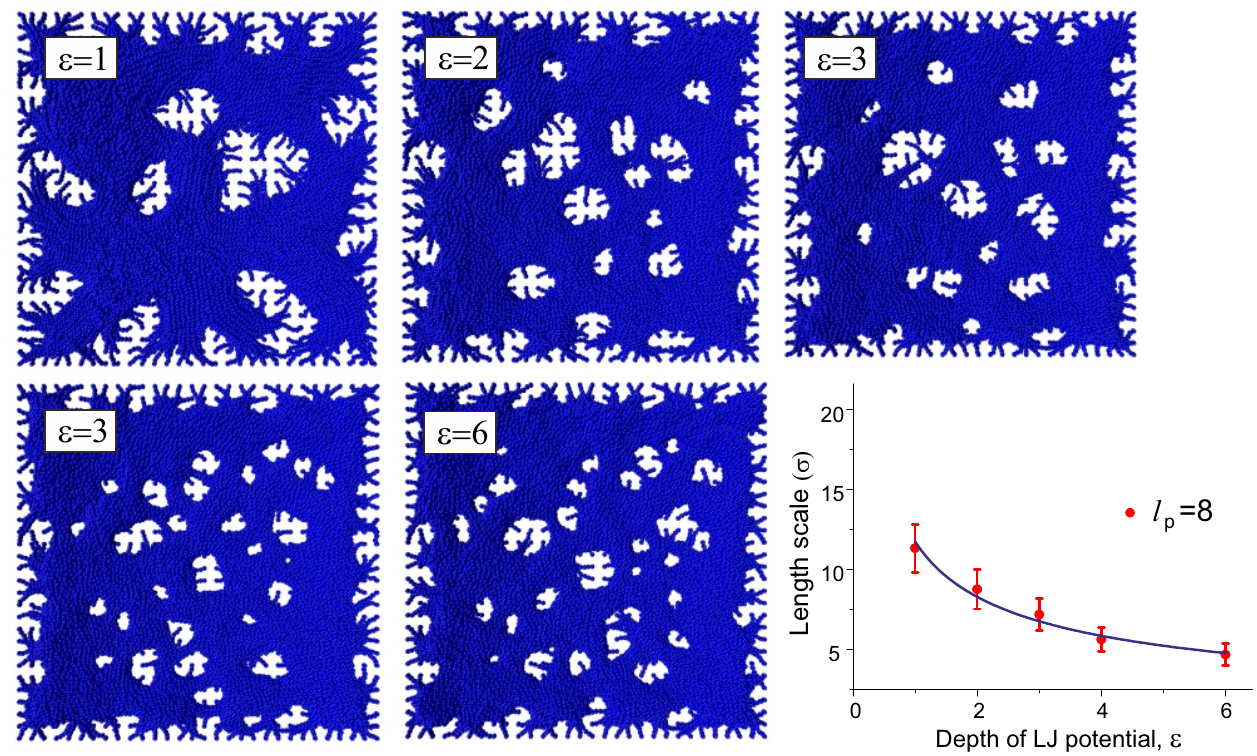}
\caption{Morphological properties of the collapsed semiflexible polymer brushes ($l_p=8$) at varying attractive potential well-depths, $\varepsilon$, labelled on each panel. The plot of the average characteristic length scale of the spinodal pattern is fitted to the square-root dependence: $\xi \sim 1/\sqrt{\varepsilon}$. }
\label{fig3}
\end{figure}

For semiflexible brushes with persistence length $l_p = 8\sigma$, there is a similar trend in morphological properties as a function of attractive potential strength -- as the potential increases, the number of chain clusters also increases and these become smaller in size, as demonstrated in fig.~\ref{fig3}. The similar quantitative analysis of the average characteristic length scale of the frozen spinodal pattern follows a similar scaling dependence $\xi \sim 1/\sqrt{\varepsilon}$ (acknowledging the large errors of its calculation and only a few data points available for the fitting). Comparing with the flexible brush ($l_p=1\sigma$) the values of this length are uniformly lower by  a constant factor of $\sim 1.33$ in this case of semiflexible chains.

\begin{figure} [t]
\centering
\includegraphics[width=.9\textwidth]{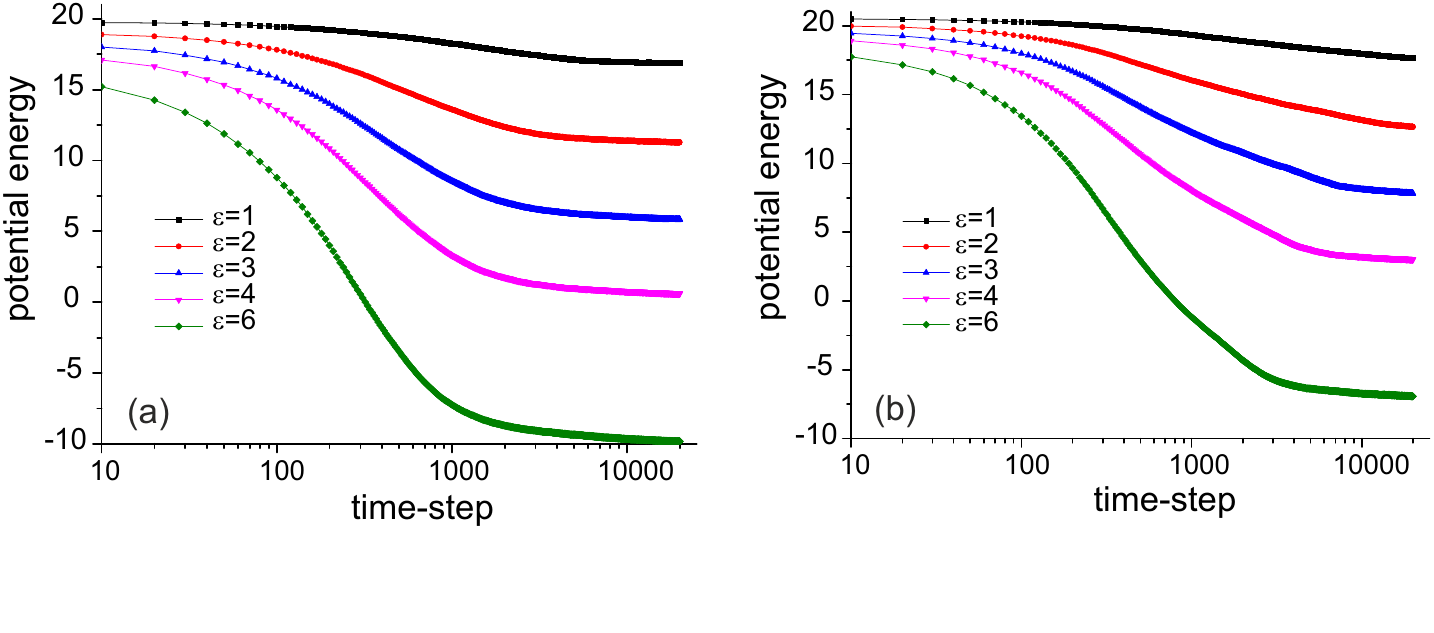}
\caption{ Short-time evolution of the potential energy of polymer brushes as a function of time. (a) Flexible chains, $l_p=1$;  (b)  semiflexible chains, $l_p=8$. Different curves correspond to increasing values of effective attraction measured by $\varepsilon$, labelled on plots. }
\label{fig4}
\end{figure}

Figure~\ref{fig4} shows the `early time' relaxation of the potential energy of each brush after the instantaneous quench at $t=0$. The potential energy experiences a sudden drop at short times, when large gains could be made by all attracting particles coming closer together; the amplitude of this drop is obviously proportional to the depth of the attractive LJ potential well. After the system becomes compact (the density plots after $t \approx 10$ kts do not have much of the further variation, fig.~\ref{fig5} below), the approach to equilibrium becomes very slow. This is due to structural changes in chain configurations in the dense state.  All structures were equilibrated for 26 million time steps and for the most of this long time the total potential energy of the 900 chains in our brushes did not vary at all.

Figure~\ref{fig5} presents the density profiles of our brushes in different conditions and at different times of their collapse, as always, comparing the flexible and the semiflexible chain cases. The first interesting result is for the flexible brush in good solvent: the density $\rho(z)$ is parabolic to a good approximation, as the fitted line in fig.~\ref{fig5}(a) indicates --  therefore our results are in full compliance with the classical theory of brushes~\cite{milner_1988}, according to which  $\rho(z) = w^{-1}(A(h) - Bz^2)$ and the effective potential acting on the chains is  $U(z)=-w\rho(z)$, with $w$  the excluded volume parameter and $h$  the brush height.  This also confirms that our initial brush is equilibrated. In contrast, the brush of semiflexible chains in  fig.~\ref{fig5}(b) is much more extended and its density does not fit a parabolic function in any approximation.

The second fact one can notice from the density $\rho(z)$ evolution on brush collapse is the fact that there is no further change in density past a time $t \sim 50$ kts (compared with the $t = 26$ Mts in the plots). This corresponds to a completed relaxation of the potential energy of the brush, cf. fig.~\ref{fig4} -- one has to assume that the further re-arrangements of the microstructure in the dense system are mainly entropy-driven.

It is evident from both the potential energy relaxation, as well as the density profile evolution, that the time that it takes for a chain to collapse is a function of the attractive potential strength. We declared the polymer brush `collapsed' (and recorded the corresponding time it took) when the density profile $\rho(z)$ have reached the shape that have no longer evolved with time (cf. fig.~\ref{fig5}). Hence we found that the stronger the attractive potential (deeper the quench), the faster this collapse takes place: see fig.~\ref{fig6}. The lines in the plot are a guide to an eye fitted dependence of $\tau \sim 1/\varepsilon$. At the same time, the semiflexible brush (more extended in good solvent) is much slower to reach the fully collapsed state on the substrate.

\begin{figure} [t]
\centering
\includegraphics[width=.9\textwidth]{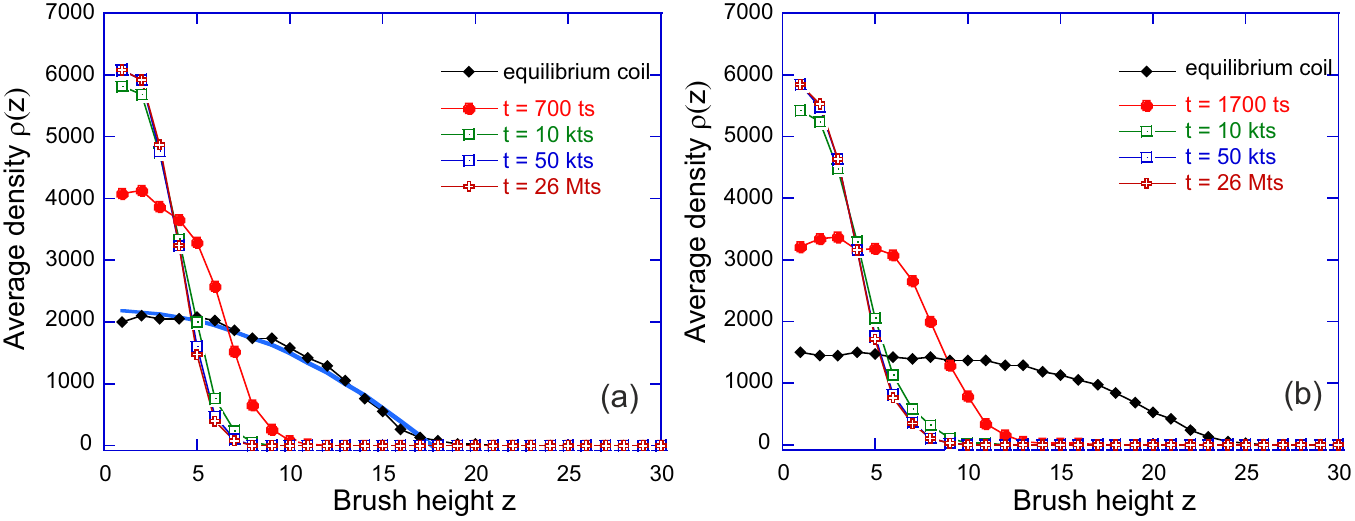}
\caption{ The evolution of density $\rho(z)$, calculated as the average number of monomers in a given $xy$-plane, at various time-points of brush collapse at $\varepsilon = 6$: (a) flexible chains, $l_p=1$;  (b)  semiflexible chains, $l_p=8$. The initial density profile in good solvent (`equilibrium coil') shows a good approximation to the classical parabolic profile in the flexible brush~\cite{milner_1988}, fitted with a solid blue line; the corresponding semiflexible chains are much more extended near the grafting plane and only show the expected drop in density from about half-way along the brush height. }
\label{fig5}
\end{figure}

\begin{figure}
\centering
\includegraphics[width=.45\textwidth]{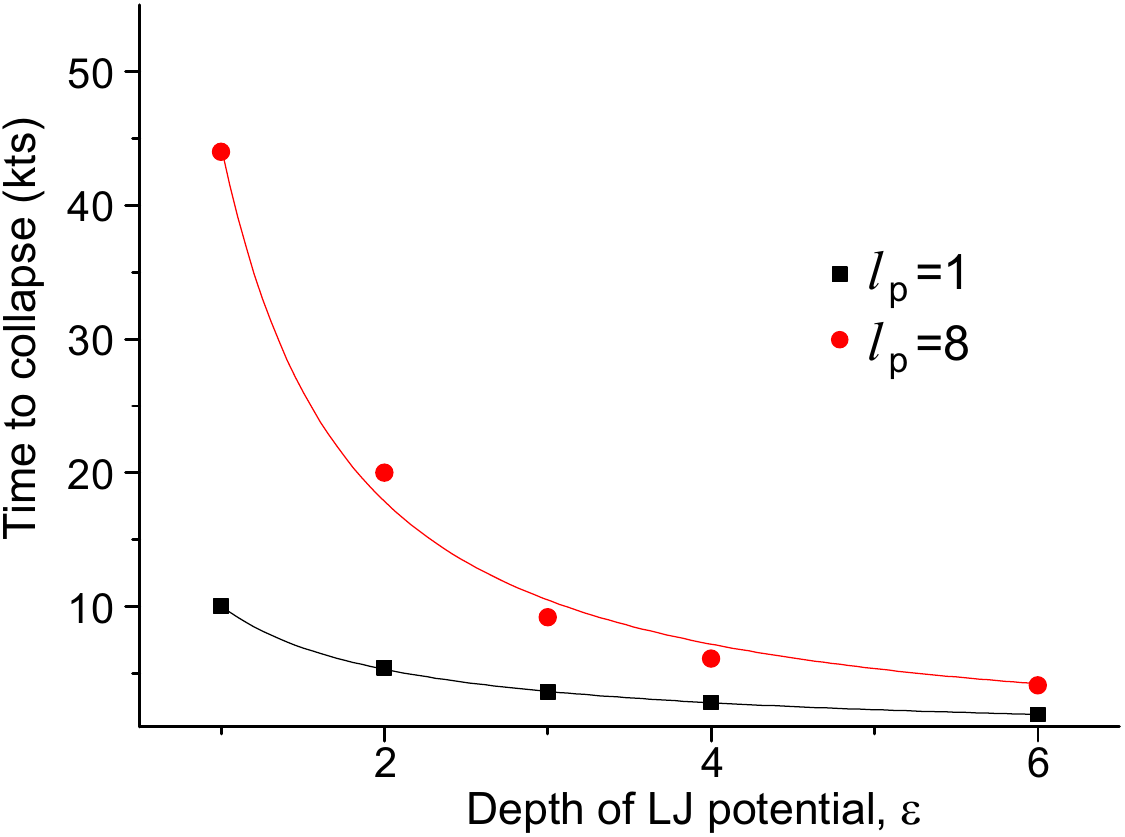}
\caption{ Time to collapse as a function of potential well depth for flexible and semiflexible polymer brushes. The lines are guide to an eye fits with an inverse function $\tau = A/\varepsilon$.}
\label{fig6}
\end{figure}

\section{Discussion} \label{section5}

There is one example in the literature, where a competition between the attractive forces between all chain segments and the constraint of remaining uniformly grafted on a flat surface has been studied theoretically~\cite{benetatos_2013}.  The direct analogy with our simulation work is difficult to establish, because the authors there have only consider the so-called directed polymers -- the chains never allowed to fold back upon themselves by either the constraint of a fixed height of the brush (the second plane fixing the ends of the chains as well), or by having a very high persistence length. Both of these limits have only marginal correspondence to our system, which is collapsing into a highly folded dense melt (our semiflexible chain case might be the closest to what has been studied in~\cite{benetatos_2013}. Nevertheless, the fundamental physics of a spatial instability developing in the $xy$-plane in order to resolve the mentioned competition has to be the same in both works, which is why we reflect on this comparison.

In essence, the final answer of that theoretical study \cite{benetatos_2013} is a prediction that an initially homogeneous areal density $\rho(x,y)$ will develop an instability when the following condition is first met (written in an approximate form, in the limit of the range of the effective attractive LJ interaction less than $\sqrt{L^3/l_p}$, following  \cite{benetatos_2013}):
\begin{equation}\label{beneta}
\varepsilon\frac{\sigma}{a^2} \geq \frac{q^4}{q^2-3+4 e^{-q^2/2} - e^{-q^2}},
\end{equation}
where $a$ is the separation of the grafting points in the plane (i.e. $1/a^2$ is the grafting density), and the non-dimensional parameter $q = \textbf{k}\sqrt{L^3/l_p}$ is proportional to the Fourier wave vector of the polymer density in the $xy$-plane. The minimum of the universal function in the left-hand side of (\ref{beneta}) is at $q_0 \approx 2$, which means the spinodal-decomposition instability will first occur with a length scale $\xi_0 \sim 1/k_0 \approx \sqrt{L^3/l_p}$  when the strength of attraction potential $\varepsilon$ exceeds a critical value $\sim 10a^2/\sigma$.

As mentioned earlier, we cannot hope to achieve quantitative agreement with this theory: in our systems the chains can fold into a densely packed state where the approximations made to carry out their calculations fail. Besides, we only examine the final state of the arrested spinodal decomposition -- not the initial point of instability.  Nevertheless, we do see that the characteristic length scale of the inhomogeneous structure  $\xi$ is smaller in a system with a larger persistence length (fig.~\ref{fig3}), and the dependence on the quenching depth is also in full correspondence with the physics of spinodal decomposition away from the critical point. We feel that these physical ideas and considerations should initiate a more focused theoretical and simulation effort to establish the laws and the morphology of polymer brushes collapsed in poor solvent -- now that we know their density in the plane can be highly non-uniform.
We also believe that this analysis can be used to finely tune polymer brushes for experimental purposes to regulate the surface properties of materials. Our results can facilitate the physical understanding of polymer brushes in experimental applications such as directed drug-delivery systems, biosensors and polymer brush modified membranes for protein analysis.

\subsection*{Acknowledgments}
This work was funded by the Osk. Huttunen Foundation (Finland), and the Cambridge Theory of Condensed Matter Grant from EPSRC. Simulations were performed using the Darwin supercomputer of the University of Cambridge High Performance Computing Service (http://www.hpc.cam.ac.uk/), provided by Dell Inc. using Strategic Research Infrastructure funding from the Higher Education Funding Council for England.

\providecommand*\mcitethebibliography{\thebibliography}
\csname @ifundefined\endcsname{endmcitethebibliography}
  {\let\endmcitethebibliography\endthebibliography}{}

\end{document}